\documentclass[twocolumn]{emulateapj}

\usepackage{natbib}
\usepackage{float}
\usepackage{amsmath}
\restylefloat{table}
\usepackage{graphicx}
\usepackage{hyperref}
\usepackage{listings}
\usepackage{textcomp}
\usepackage{upquote}
\usepackage{mathtools}

\usepackage{textcomp}   
\usepackage[nothing]{algorithm} 
\usepackage{color} 
\definecolor{codegreen}{rgb}{0,0.6,0} 
\definecolor{codeorange}{rgb}{1,0.6,0} 
\definecolor{codegray}{rgb}{0.5,0.5,0.5} 
\definecolor{codepurple}{rgb}{0.58,0,0.82} 
\definecolor{backcolour}{rgb}{0.95,0.95,0.92} 
\usepackage{listings} 
\lstnewenvironment{python}[1][]{ 
\lstset{   
language=python,   
basicstyle=\ttfamily\small,     
keywordstyle=\color{codeorange},   
stringstyle=\color{codegreen},   
showstringspaces=false,   
emph={class, pass, in, for, while, if, is, elif, else, not, and, or, from, import, as, True, False, None, 
def, exec, break, continue, return},   
emphstyle=\color{codeorange}\bfseries,   
emph={[2]print, setattr, getattr},   
emphstyle=[2]\color{codepurple},   
emph={[3]self, cls},   
emphstyle=[3]\color{magenta},  
 upquote=true,   
 morecomment=[s]{"""}{"""},   
 commentstyle=\color{codegray}\slshape,   
 tabsize=4 
 }}{}

\usepackage{fmtcount}
\usepackage{verbatim}

\usepackage{algorithm}
\usepackage[noend]{algpseudocode}
\usepackage{epstopdf}
\usepackage{epsfig}

\newcommand{\quotes}[1]{``#1''}

\begin{document}
\bibliographystyle{unsrt}
\lstset{language=Python,upquote=true, breaklines=true, basicstyle=\small}

\title{FATS: Feature Analysis for Time Series}
\author{Isadora Nun$^{1}$,  Pavlos Protopapas$^{1,2}$, \\ Brandon Sim$^{1}$, Ming Zhu$^{1}$, Rahul Dave$^{1}$, Nicolas Castro$^{3}$,  Karim Pichara$^{3,4,1}$ }
\affil{\altaffilmark{1}Institute for Applied Computational Science, Harvard University, Cambridge, MA, USA}
\affil{\altaffilmark{2}Harvard-Smithsonian Center for Astrophysics, Cambridge, MA, USA}
\affil{\altaffilmark{3}Computer Science Department, Pontificia Universidad Cat\'olica de Chile, Santiago, Chile}
\affil{\altaffilmark{4}The Millennium Institute of Astrophysics}

\begin{abstract}
In this paper, we present the FATS (Feature Analysis for Time Series) library. FATS is a Python library which facilitates and standardizes feature extraction for time series data. In particular, we focus on one application: feature extraction for astronomical light curve data, although the library is generalizable for other uses. We detail the methods and features implemented for light curve analysis, and present examples for its usage.
\end{abstract}

\keywords{methods: data analysis -- methods: statistical -- stars: statistics -- stars: variables: general -- catalogs}

\section{Introduction}

A time series is a sequence of observations, or data points, that is arranged based on the times of their occurrence. The hourly measurement of wind speeds in meteorology, the minute by minute recording of electrical activity along the scalp in electroencephalography, and the weekly changes of stock prices in finances are just some examples of time series, among many others. Some of the following properties may be observed in time series data \citep{Falk2012}:
\begin{itemize}
\item the data is not generated independently
\item their dispersion varies in time
\item they are often governed by a trend and/or have cyclic components
\end{itemize}
The study and analysis of time series can have multiple ends: to gain a better understanding of the mechanism generating the data, to predict future outcomes and behaviors, to classify and characterize events, or more.

In time domain astronomy, data gathered from telescopes is usually represented in the form of light curves, which are time series that show the brightness variation of an object through a period of time. Based on the variability characteristics of the light curves, celestial objects can be classified into different groups (quasars, long period variables, eclipsing binaries, etc., as shown in figure \ref{fig:curvas_ejemplos} \citep{Nun2014a}) and consequently can be studied in depth independently.


\begin{figure}[h]
\centering
\includegraphics[width=8.5cm]{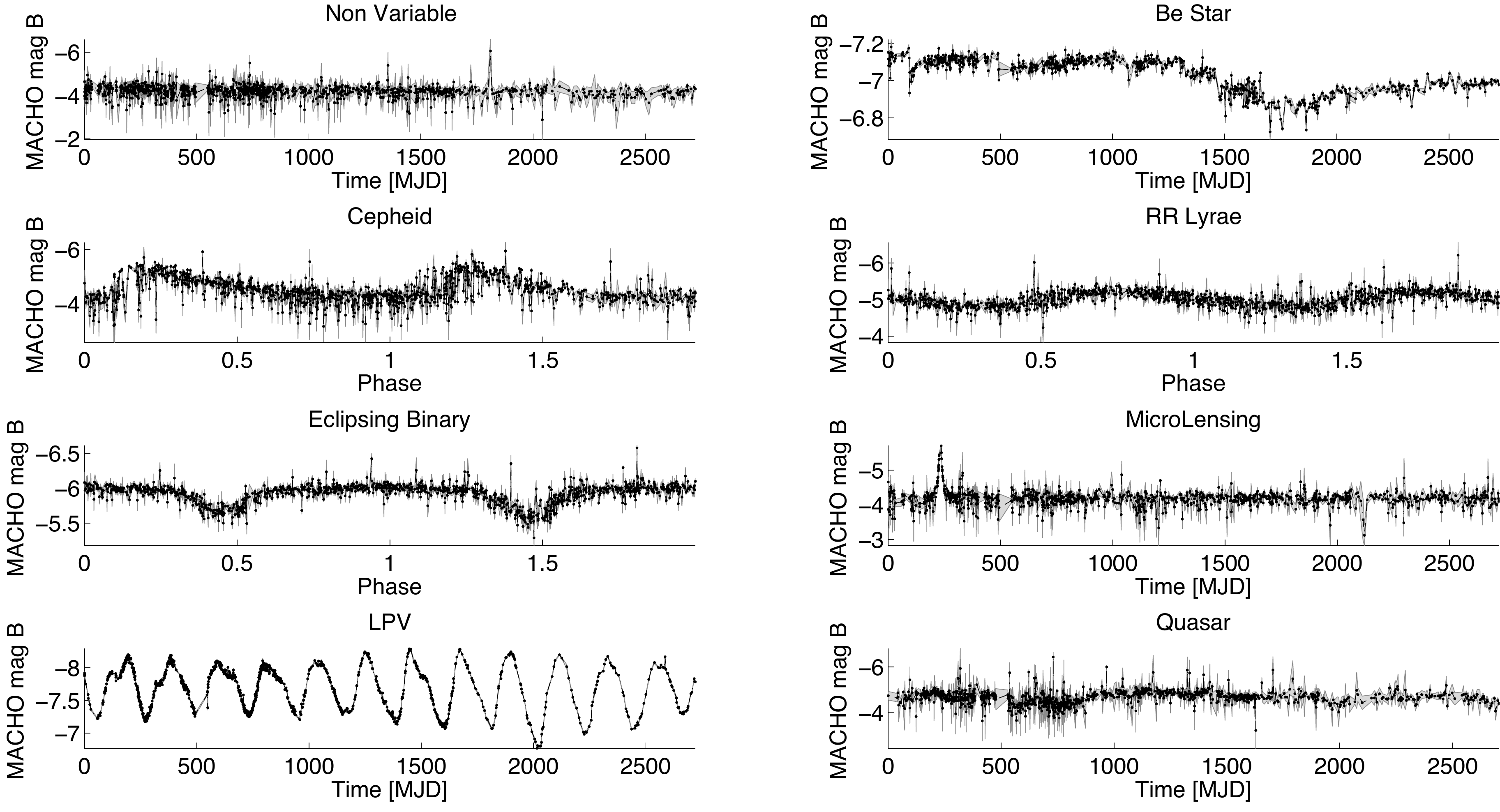}
\caption{Example light curves of each class in the MACHO training set.}
\label{fig:curvas_ejemplos}
\end{figure}

Classification of data into groups can be performed in several ways given light curve data: primarily, existing methods found in the literature use machine learning algorithms that group light curves into categories through feature extraction from the light curve data. These light curve features, the topic of this work, are numerical or categorical properties of the light curves which can be used to characterize and distinguish the different variability classes. Features can range from basic statistical properties such as the mean or the standard deviation to more complex time series characteristics such as the autocorrelation function. These features should ideally be informative and discriminative, thus allowing for machine learning or other algorithms to use them to distinguish between classes of light curves.

In this document, we present a library which allows for the fast and efficient calculation of a compilation of many existing light curve features. The main goal is to create a collaborative and open tool where users can characterize or analyze an astronomical photometric database while also contributing to the library by adding new features. This arXiv document as well as the authors listed will be updated insofar as new features are created. The reader can also visit the library website \url{http://isadoranun.github.io/tsfeat/FeaturesDocumentation.html} for more information and a better visualization of the tool. It is important to highlight that this library is not necessarily restricted to the astronomical domain and can also be applied to any kind of time series data.

Our vision is to be capable of analyzing and comparing light curves from any available astronomical catalog in a standard and universal way. This would facilitate and make more efficient tasks such as modeling, classification, data cleaning, outlier detection, and data analysis in general. Consequently, when studying light curves, astronomers and data analysts using our library would be able to compare and match different features in a standardized way. In order to achieve this goal, the library should be run and features generated for every existent survey (MACHO, EROS, OGLE, Catalina, Pan-STARRS, etc.), as well as for future surveys (LSST), and the results shared openly, as is this library.

In the remainder of this document, we provide an overview of the features developed so far and explain how users can contribute to the library.

\section{Light Curve Features}
\label{sec:features}
The next section details the features that we have developed in order to represent light curves. For each feature, we also describe a benchmark test performed in order to test the feature's correctness.

Each feature was also tested to ensure invarability to unequal sampling. Because telescope observations are not always taken at uniformly spaced intervals, the light curve features should be invariant to this nonuniformity. These tests are explained in section \ref{sec:invariance}.

\subsection{Mean }

Mean magnitude
\begin{equation}
\bar{m} = \frac{\sum_{i=1}^N {m_i}}{N}
\end{equation}

\subsection{Standard deviation }

The standard deviation, $\sigma$, of the sample is defined as:

\begin{equation}
\sigma=\sqrt{\frac{1}{N-1}\sum_{i} (m_{i}-\bar{m})^2}
\end{equation}


\subsection{Mean variance $\frac{\sigma}{\bar{m}}$ }

This is a simple variability index and is defined as the ratio of the standard deviation, $\sigma$, to the mean magnitude, $\bar{m}$. If a light curve has strong variability, $\frac{\sigma}{\bar{m}}$ of the light curve is generally large.

For a uniform distribution from 0 to 1, the mean is equal to 0.5 and the variance is equal to $1/12$, thus the mean-variance should take a value close to 0.577.

\subsection{Median buffer range percentage (MedianBRP) \citep{Richards2011}}

Fraction of photometric points within amplitude/10 of the median magnitude.

\subsection{Range of a cumulative sum $R_{cs}$ \citep{Kim2011} }
\label{sec:rcs}
$R_{cs}$ is the range of a cumulative sum \citep{Ellaway1978} of each light curve and is defined as:

\begin{align*}
R_{cs} &= \max S - \min S\\
S &= \frac{1}{N \sigma} \sum_{i=1}^l \left( m_i - \bar{m} \right)
\end{align*}

for $l=1,2, \ldots, N$. 

$R_{cs}$ should take a value close to zero for any symmetric distribution.

\subsection{Period (Lomb-Scargle) \citep{Kim2011}}

The Lomb-Scargle (L-S) algorithm\citep {scargle1982} is a common tool used in time series for period finding and frequency analysis. It is usually chosen over the Discrete Fourier Transform (DFT) because it can handle unevenly spaced data points. In this peridiogram time series are decomposed into a linear combination of sine waves of the form $y = a \cos \omega t + b \sin\omega t$. By doing this fit, the data is transformed from the time domain to the frequency domain. 
The L-S peridiogram,  is defined as:

\begin{multline*}
P(\omega) = \frac{1}{2\sigma^2}\left\{\frac{\left[\sum^N_{n=1}(m_n - \bar{m})\cos\left[\omega(t_n-\tau)\right]\right]^2}{\sum^N_{n=1}\cos^2\left[\omega(t_n-\tau)\right]} \right.\\
\left. + \frac{\left[\sum^N_{n=1}(m_n - \bar{m})\sin\left[\omega(t_n-\tau)\right]\right]^2}{\sum^N_{n=1}\sin^2\left[\omega(t_n-\tau)\right]}\right\}
\end{multline*}

where $\omega = 2\pi T$, $T$ being the period and  the time offset $\tau$ is defined by:
$$ \tan(2\omega \tau) = \frac{\sum^N_{n=1} \sin(2\omega t_n)}{\sum^N_{n=1} \cos(2\omega t_n)} .$$

%

On the other hand, periodic light curves can be transformed into a single light curve in which each period is mapped onto the same time axis as follows:

\begin{equation*}
t'=\left\{\frac{t-t_0}{T}\right\}
\end{equation*}

where $T$ is the period, $t_0$ is an arbitrary starting point and the symbol \{\} represents the non-integer part of the fraction. This process produces a folded light curve on an $x$-axis of folded time that ranges from 0 to 1. 

\begin{figure}[h!]
\centering
\includegraphics[width=8.5cm]{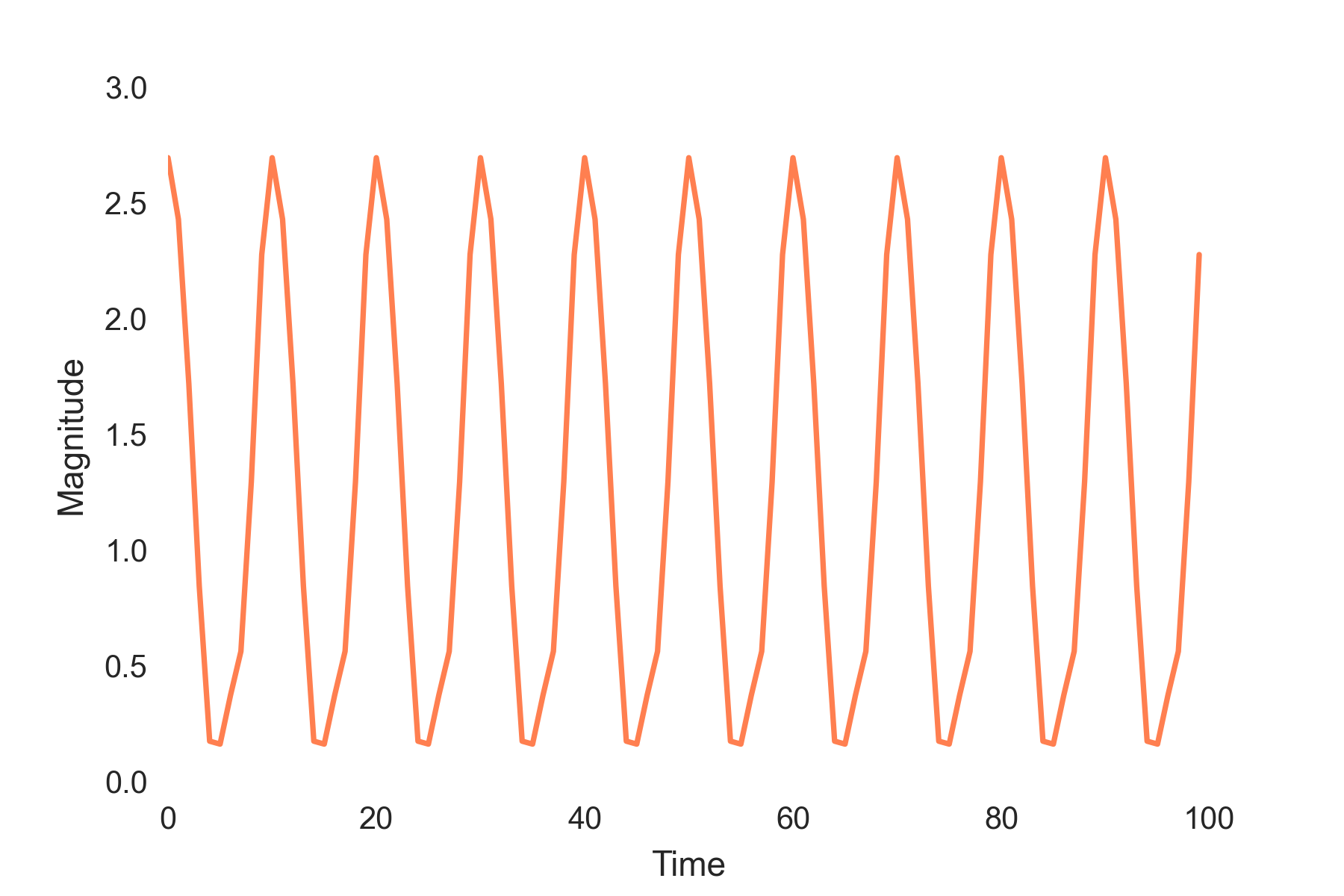}
\includegraphics[width=8.5cm]{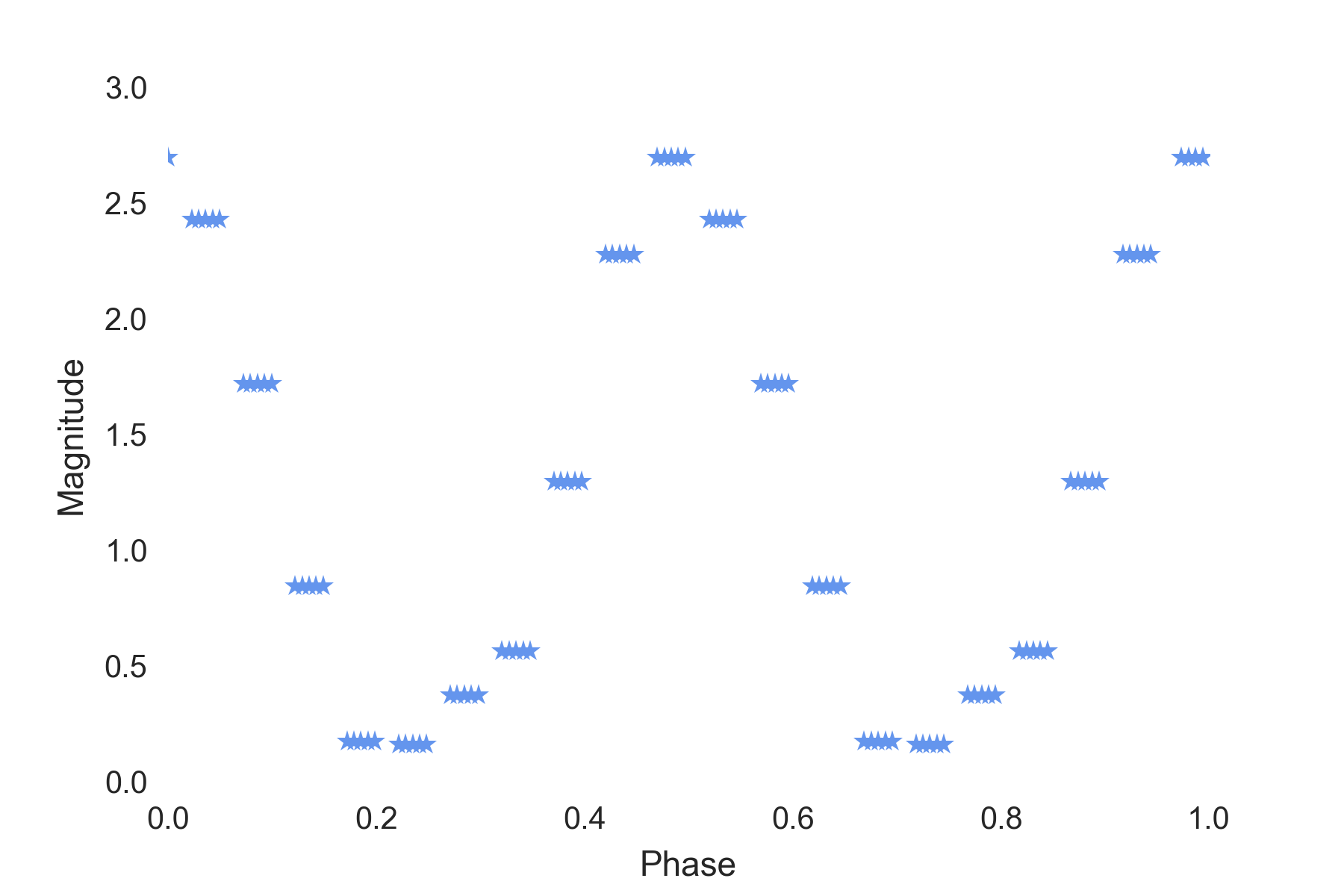}
\caption{Synthetic time series before and after folding.}
\label{fig:folded}
\end{figure}

In order to benchmark correctness of this feature, we calculate the period of a synthetic periodic time series and compare the obtained value with the real one. We also calculate the period of a known periodic light curve and fold it as shown in figure \ref{fig:folded}.

\subsection{Period fit \citep{Kim2011}}

We define period fit as the false-alarm probability of the largest periodogram value obtained with the Lomb-Scargle algorithm. Its value should be close to zero for a periodic time series.

\subsection{Range of a cumulative sum on phase-folded light curve, $\Psi_{CS}$ \citep{Kim2014}}

Range of a cumulative sum (sec \ref{sec:rcs}) applied to the phase-folded light curve (generated using the period estimated from the Lomb-Scargle method).

\subsection{Periodic features extracted from light curves using generalized Lomb–Scargle \citep{Richards2011} }

Here, we adopt a model where the time series of the photometric magnitudes of variable stars is modeled as a superposition of sines and cosines:

$$m_i(t|f_i) = a_i\sin(2\pi f_i t) + b_i\cos(2\pi f_i t) + b_{i,\circ}$$

where $a$ and $b$ are normalization constants for the sinusoids of frequency $f_i$ and $b_{i,\circ}$ is the magnitude offset \citep{Richards2011}.
To find periodic variations in the data, we fit the equation above by minimizing the sum of squares, which we denote $\chi^2$:

\begin{equation}
\chi^2 = \sum_k \frac{(d_k - m_i(t_k))^2}{\sigma_k^2}
\end{equation}

where $\sigma_k$ is the measurement uncertainty in data point $d_k$. We allow the mean to float, leading to more robust period estimates in the case where the periodic phase is not uniformly sampled; in these cases, the model light curve has a non-zero mean. This can be important when searching for periods on the order of the data span $T_{\textrm{tot}}$. Now, define

\begin{equation}
\chi^2_{\circ} = \sum_k \frac{(d_k - \mu)^2}{\sigma_k^2}
\end{equation}

where $\mu$ is the weighted mean

\begin{equation}
\mu = \frac{\sum_k d_k / \sigma_k^2}{\sum_k 1/\sigma_k^2}
\end{equation}

Then, the generalized Lomb-Scargle periodogram is:

\begin{equation}
P_f(f) = \frac{(N-1)}{2} \frac{\chi_{\circ}^2 - \chi_m^2(f)}{\chi_{\circ}^2}
\end{equation}

where $\chi_m^2(f)$ is $\chi^2$ minimized with respect to $a, b$, and $b_{\circ}$.

Following \citet{debosscher2007}, we fit each light curve with a linear term plus a harmonic sum of sinusoids:

\begin{equation}
m(t) = ct + \sum_{i=1}^{3}\sum_{j=1}^{4} m_i(t|jf_i)
\end{equation}

where each of the three test frequencies $f_i$ is allowed to have four harmonics at frequencies $f_{i,j} = jf_i$. The three test frequencies $f_i$ are found iteratively, by successfully finding and removing periodic signal producing a peak in $P_f(f)$, where $P_f(f)$ is the Lomb-Scargle periodogram as defined above.

Given a peak in $P_f(f)$, we whiten the data with respect to that frequency by fitting away a model containing that frequency as well as components with frequencies at 2, 3, and 4 times that fundamental frequency (harmonics). Then, we subtract that model from the data, update $\chi_{\circ}^2$, and recalculate $P_f(f)$ to find more periodic components.

\begin{algorithm}[H]
\caption{Periodic features extracted from light curves using generalized Lomb-Scargle}\label{harmonic-algo}
\begin{algorithmic}[1]
\Procedure{FindPeriodicComponents}{}
\For{$i\gets 1, 2, 3$}
\State Calculate Lomb-Scargle periodogram $P_f(f)$ for light curve.
\State Find peak in $P_f(f)$, subtract that model from data.
\State Update $\chi_{\circ}^2$.
\EndFor
\EndProcedure
\end{algorithmic}
\end{algorithm}

Then, the features extracted are given as an amplitude and a phase:

\begin{align}
A_{i,j} &= \sqrt{a_{i,j}^2 + b_{i,j}^2}\\
\textrm{PH}_{i,j} &= \arctan\left(\frac{b_{i,j}}{a_{i,j}}\right)
\end{align}

where $A_{i,j}$ is the amplitude of the $j$th harmonic of the $i$th frequency component and $\textrm{PH}_{i,j}$ is the phase component, which we then correct to a relative phase with respect to the phase of the first component:

\begin{equation}
\textrm{PH}'_{i,j} = \textrm{PH}_{i,j} - \textrm{PH}_{00}
\end{equation}

and remapped to $|-\pi, +\pi|$.

\subsection{Color \citep{Kim2011}}

This feature only applies to astronomical light curves and is defined as the difference between the average magnitude of two different bands observations. 

\begin{figure}[H]
\centering
\includegraphics[width=8.5cm]{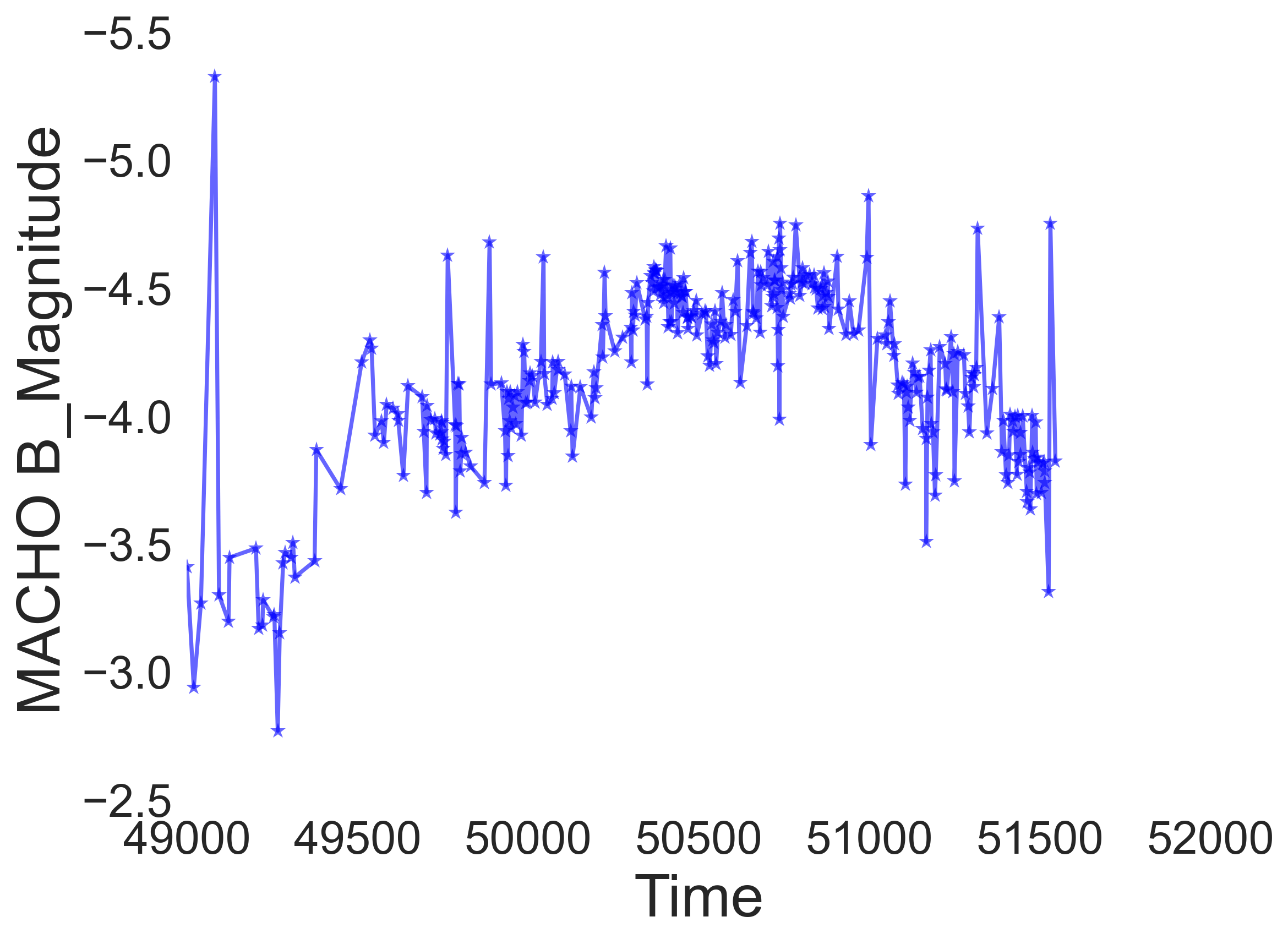}
\label{fig:curvas_ejemplos}
\end{figure}

\begin{figure}[H]
\centering
\includegraphics[width=8.5cm]{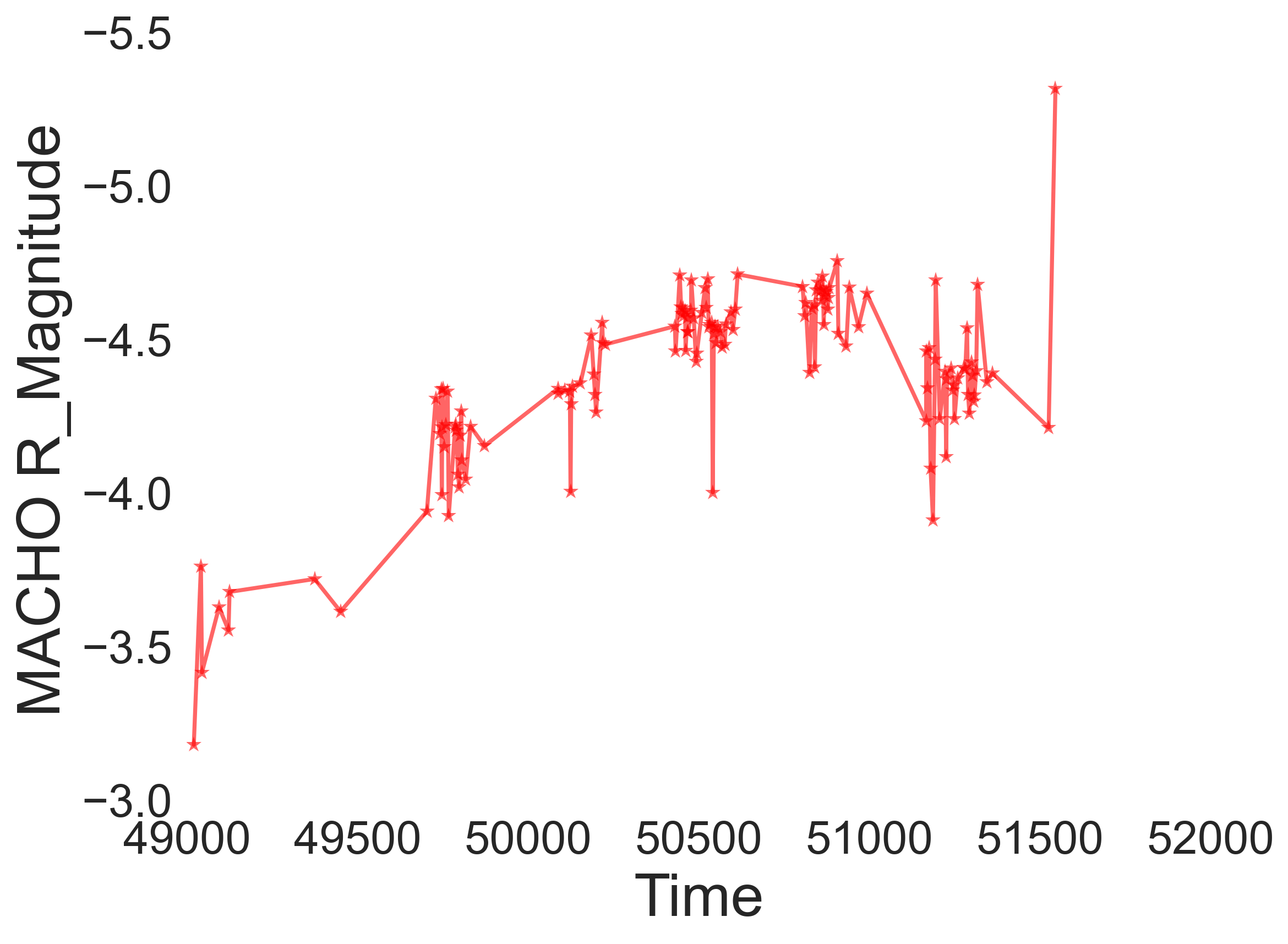}
\caption{Blue and red band observations for a particular light curve.}
\label{fig:curvas_ejemplos}
\end{figure}

\subsection{Autocorrelation function length \citep{Kim2011}}

The autocorrelation or serial correlation, is the linear dependence of a signal with itself at two points in time. In other words, it represents how similar the observations are as a function of the time lag between them. It is often used for detect non-randomness in data or to find repeating patterns.

For an observed series $m_1, m_2,\dots,m_T$ with sample mean $\bar{m}$, the sample lag$-h$ autocorrelation is given by:

$$\hat{\rho}_h = \frac{\sum_{t=h+1}^T (m_t - \bar{m})(m_{t-h}-\bar{m})}{\sum_{t=1}^T (m_t - \bar{m})^2}$$

Since the autocorrelation function of a light curve is given by a vector and we can only return one value as a feature, we define the length of the autocorrelation function as the lag value where $\hat{\rho}_h$ becomes smaller than $e^{-1}$. 

\subsection{Slotted autocorrelation function length}

In slotted autocorrelation \citep{huijse2012}, time lags are defined as intervals or slots instead of single values. The slotted autocorrelation function at a certain time lag slot is computed by averaging the cross product between samples whose time differences fall in the given slot.

$$\hat{\rho}(\tau=kh) = \frac {1}{\hat{\rho}(0)\,N_\tau}\sum_{t_i}\sum_{t_j= t_i+(k-1/2)h }^{t_i+(k+1/2)h } \bar{m}_i(t_i)\,\, \bar{m}_j(t_j) $$

Where $h$ is the slot size, $\bar{y}$ is the normalized magnitude, $\hat{\rho}(0)$ is the slotted autocorrelation for the first lag, and $N_\tau$ is the number of pairs that fall in the given slot.  Again, we define the length of the slotted autocorrelation function as  the lag value where the $\hat{\rho}(\tau=kh)$  becomes smaller than $e^{-1}$. 

In order to check the validity of this feature we calculated the slotted autocorrelation function for a normal distribution with $h=1$ and compared it with the autocorrelation function. These two should be equivalent in this case where the time intervals are constant.

\begin{figure}[H]
\centering
\includegraphics[width=8.5cm]{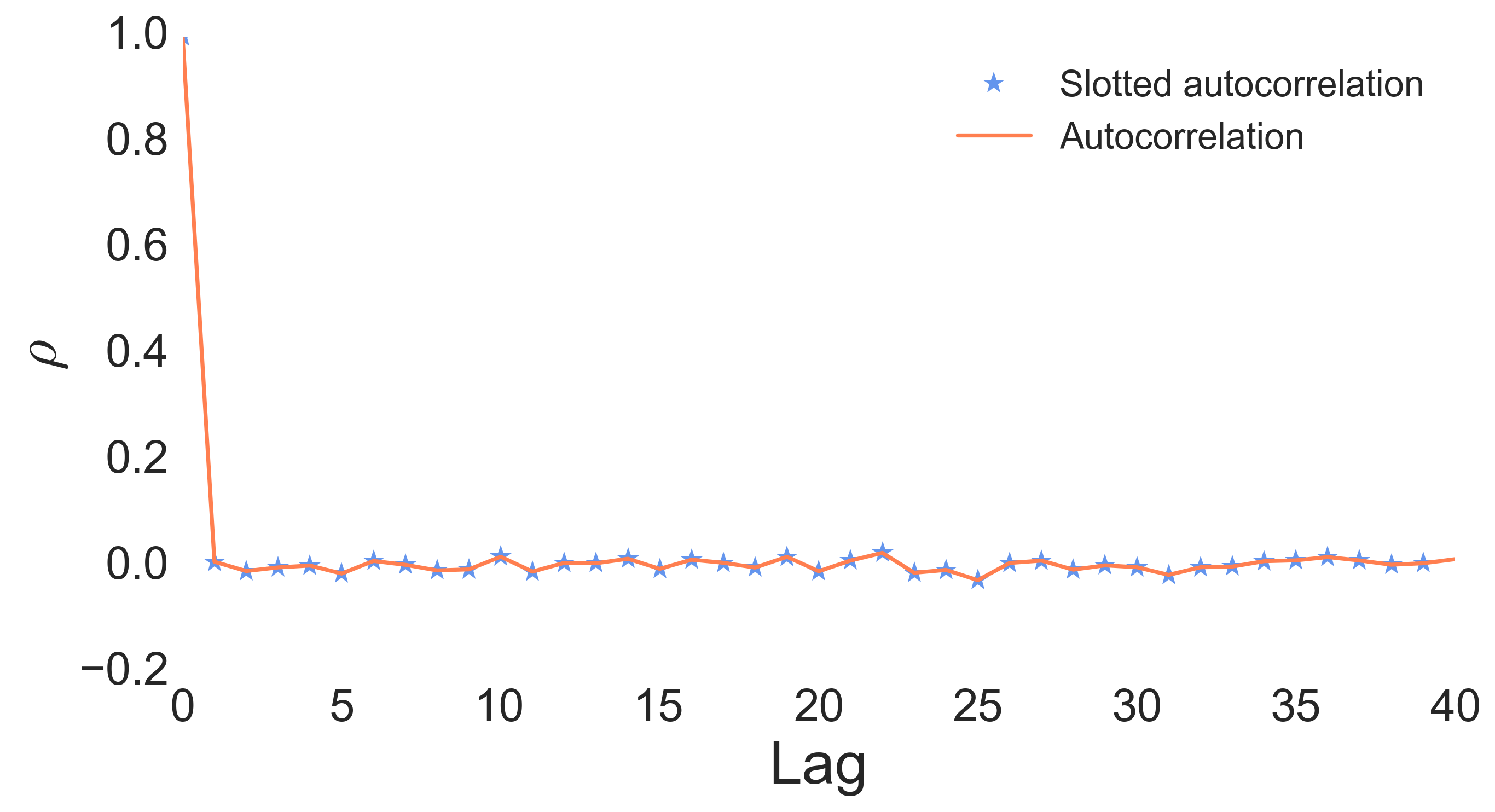}
\caption{Autocorrelation function and Slotted autocorrelation function for a normal distributed time series with $T=1$.}
\label{fig:curvas_ejemplos}
\end{figure}

\subsection{Stetson K, Stetson K\_AC, Stetson J and Stetson L \citep{Kim2011}}

These four features are based on the Welch/Stetson variability index $I$ \citep{Stetson1996} defined by the equation:
$$ I = \sqrt{\frac{1}{n(n-1)}} \sum_{i=1}^n {\left( \frac{b_i-\hat{b}}{\sigma_{b,i}}\right) \left( \frac{v_i - \hat{v}}{\sigma_{v,i}} \right)} $$

where $b_i$ and $v_i$ are the apparent magnitudes obtained for the candidate star in two observations closely spaced in time on some occasion $i$, $\sigma_{b,i}$ and $\sigma_{v,i}$ are the standard errors of those magnitudes, $\hat{b}$ and $\hat{v}$ are the weighted mean magnitudes in the two filters, and $n$ is the number of observation pairs.

Since a given frame pair may include data from two filters which did not have equal numbers of observations overall, the \quotes{relative error} is calculated as follows:

$$ \delta = \sqrt{\frac{n}{n-1}} \frac{v-\hat{v}}{\sigma_v} $$

allowing all residuals to be compared on an equal basis. 

\subsubsection{Stetson K}

Stetson K is a robust kurtosis measure:
$$ \frac{1/N \sum_{i=1}^N |\delta_i|}{\sqrt{1/N \sum_{i=1}^N \delta_i^2}}$$

where the index $i$ runs over all $N$ observations available for the star without regard to pairing. For a Gaussian magnitude distribution K should take a value close to $\sqrt{2/\pi} = 0.798$. The proof can be found in \citep{Stetson1996}.

\subsubsection{Stetson K\_AC}

Stetson K applied to the slotted autocorrelation function of the light curve.

\subsubsection{Stetson J}

Stetson J is a robust version of the variability index. It is calculated based on two simultaneous light curves of a same star and is defined as:

$$ J =  \sum_{k=1}^n  sgn(P_k) \sqrt{|P_k|}$$

with $P_k = \delta_{i_k} \delta_{j_k} $

For a Gaussian magnitude distribution, $J$ should take a value close to zero.

\subsubsection{Stetson L}

Stetson L variability index describes the synchronous variability of different bands and is defined as:
$$ L = \frac{JK}{0.798} $$

Again, for a Gaussian magnitude distribution, L should take a value close to zero.

\subsection{Variability index $\eta^e$ \citep{Kim2014}}

Variability index $\eta$  \citep{von1941} is the ratio of the mean of the square of successive differences to the variance of data points. The index was originally proposed to check whether the successive data points are independent or not. In other words, the index was developed to check if any trends exist in the data. It is defined as:
$$\eta=\frac{1}{\left(N-1 \right)\sigma^2}\sum_{i=1}^{N-1} \left( m_{i+1}-m_i \right)^2 $$

Although $\eta$ is a powerful index for quantifying variability characteristics of a time series, it does not take into account unequal sampling. Thus we use $\eta^e$, which is defined as:

$$ \eta^e = \bar{w} \, \left( t_{N-1} - t_1 \right)^2 \frac{\sum_{i=1}^{N-1} w_i \left(m_{i+1} - m_i \right)^2}{\sigma^2 \sum_{i=1}^{N-1} w_i} $$

$$ w_i = \frac{1}{\left( t_{i+1} - t_i \right)^2} $$

The variability index should take a value close to two for uncorrelated normally distributed  time series with mean equals to zero and  large N. 

\subsection{Variability index $\eta_{color}$ \citep{Kim2014}}

$\eta^e$ index calculated from the color light curve (color as a function of time).

\subsection{$\Psi_\eta$}

$\eta^e$ index calculated from the folded light curve.

\subsection{Small Kurtosis \citep{Richards2011}}

Small sample kurtosis of the magnitudes:
\begin{align*}
\kappa &= \frac{N \left( N+1 \right)}{\left( N-1 \right) \left( N-2 \right) \left( N-3 \right)} \sum_{i=1}^N \left( \frac{m_i-\hat{m}}{\sigma} \right)^4 \\
&- \frac{3\left( N-1 \right)^2}{\left( N-2 \right) \left( N-3 \right)}
\end{align*}

For a normal distribution, the small kurtosis should be zero.

\subsection{Skewness ($\gamma_1$) \citep{Richards2011}}

The skewness of a sample is defined as follow:
$$ \gamma_1= \frac{N}{\left(N-1\right)\left(N-2\right)} \sum_{i=1}^N \left( \frac{m_i-\hat{m}}{\sigma}\right)^3 $$

For a normal distribution it should be equal to zero.

\subsection{Median absolute deviation (MAD) \citep{Richards2011}}

The median absolute deviation is defined as the median discrepancy of the data from the median data:

$$\textrm{MAD} = \textrm{median}\left( |\textrm{m} - \textrm{median}(\textrm{m})|\right) $$

It should take a value close to 0.675 for a normal distribution. This can be proven by using the interquartile ranges of a normal distribution.

\subsection{Amplitude \citep{Richards2011}}

The amplitude is defined as the half of the difference between the median of the maximum 5\% and the median of the minimum 5\% magnitudes.
For a sequence of numbers from 0 to 1000 the amplitude should be equal to 475.5.

\subsection{Percent amplitude \citep{Richards2011}}

Largest percentage difference between either the max or min magnitude and the median.

\subsection{Con \citep{Kim2011}}

Index introduced for the selection of variable stars from the OGLE database \citep{wozniak2000}. To calculate Con, we count the number of three consecutive data points that are brighter or fainter than $2\sigma$ and normalize the number by $N-2$. 

For a normal distribution and by considering just one star, Con should take values close to 0.045 (area under a Gaussian curve within two sigma).

\subsection{Anderson-Darling test \citep{Kim2008}}

The Anderson-Darling test is a statistical test of whether a given sample of data is drawn from a given probability distribution. When applied to testing  if a normal distribution adequately describes a set of data, it is one of the most powerful statistical tools for detecting most departures from normality.

For a normal distribution the Anderson-Darling statistic should take values close to 0.25.

\subsection{Linear trend \citep{Richards2011}}

Slope of a linear fit to the light curve.

\subsection{Max slope \citep{Richards2011}}

Maximum absolute magnitude slope between two consecutive observations.

\subsection{Beyond 1 std \citep{Richards2011}}

Percentage of points beyond one standard deviation from the weighted (by photometric errors) mean.

For a normal distribution, it should take a value close to 0.32 (area under a Gaussian curve within one sigma).

\subsection{Pair slope trend \citep{Richards2011}}

Considering the last 30 (time-sorted) measurements of source magnitude, the fraction of increasing first differences minus the fraction of decreasing first differences.

\subsection{Flux percentile ratio mid20, mid 35, mid 50, mid 65 and mid 80 \citep{Richards2011}}

In order to characterize the sorted magnitudes distribution we use percentiles. If $F_{5,95}$ is the difference between $95\%$ and $5\%$ magnitude values, we calculate the following:
\begin{itemize}
\item flux\_percentile\_ratio\_mid20: ratio $F_{40,60}/F_{5,95}$ 
\item flux\_percentile\_ratio\_mid35: ratio $F_{32.5,67.5}/F_{5,95}$ 
\item flux\_percentile\_ratio\_mid50: ratio $F_{25,75}/F_{5,95}$
\item flux\_percentile\_ratio\_mid65: ratio $F_{17.5,82.5}/F_{5,95}$ 
\item flux\_percentile\_ratio\_mid80: ratio $F_{10,90}/F_{5,95}$
\end{itemize}

For the first feature for example, in the case of a normal distribution, this is equivalent to calculate $\frac{erf^{-1}(2 \cdot 0.6-1)-erf^{-1}(2 \cdot 0.4-1)}{erf^{-1}(2 \cdot 0.95-1)-erf^{-1}(2 \cdot 0.05-1)}$. So, the expected values for each of the flux percentile features are:
\begin{itemize}
\item flux\_percentile\_ratio\_mid20 = 0.154
\item flux\_percentile\_ratio\_mid35 = 0.275 
\item flux\_percentile\_ratio\_mid50 = 0.410
\item flux\_percentile\_ratio\_mid65 = 0.568
\item flux\_percentile\_ratio\_mid80 = 0.779
\end{itemize}

\subsection{Percent difference flux percentile \citep{Richards2011}}

Ratio of $F_{5,95}$ over the median magnitude.

\subsection{$Q_{3-1}$ \citep{Kim2014}}

$Q_{3-1}$ is the difference between the third quartile, $Q_3$, and the first quartile, $Q_1$, of a raw light curve. $Q_1$ is a split between the lowest $25\%$ and the highest $75\%$ of data. $Q_3$ is a split between the lowest $75\%$ and the highest $25\%$ of data.

\subsection{$Q_{3-1|B-R}$ \citep{Kim2014}}

$Q_{3-1}$ applied to the difference between both bands of a light curve (B-R).

\subsection{CAR features \citep{Pichara2012}}

In order to model the irregular sampled times series we use CAR(1) \citep{brockwell2002}, a continuous time auto regressive model. CAR(1) process has three parameters, it provides a natural and consistent way of estimating a characteristic time scale and variance of light curves. CAR(1) process is described by the following stochastic differential equation:

\begin{align*}
dX(t) &= - \frac{1}{\tau} X(t)dt + \sigma_C \sqrt{dt} \epsilon(t) + bdt\\
&\textrm{for } \tau, \sigma_C, t \geq 0
\end{align*}

where the mean value of the light curve $X(t)$ is $b\tau$ and the variance is $\frac{\tau\sigma_C^2}{2}$. $\tau$ is the relaxation time of the process $X(T)$, it can be interpreted as describing the variability amplitude of the time series. $\sigma_C$ can be interpreted as describing the variability of the time series on time scales shorter than $\tau$. $\epsilon(t)$ is a white noise process with zero mean and variance equal to one. The likelihood function of a CAR(1) model for a light curve with observations $x - \{x_1, \dots, x_n\}$ observed at times $\{t_1, \dots, t_n\}$ with measurements error variances $\{\delta_1^2, \dots, \delta_n^2\}$ is:

$$ p \left( x|b,\sigma_C,\tau \right) = \prod_{i=1}^n \frac{1}{[2 \pi \left( \Omega_i + \delta_i^2 \right)]^{1/2}} exp \{ -\frac{1}{2} \frac{\left( \hat{x}_i - x^*_i \right)^2}{\Omega_i + \delta^2_i} \} $$
$$ x_i^* = x_i - b\tau$$
$$ \hat{x}_0 = 0 $$
$$ \Omega_0 = \frac{\tau \sigma^2_C}{2} $$
$$ \hat{x}_i = a_i\hat{x}_{i-1} + \frac{a_i \Omega_{i-1}}{\Omega_{i-1} + \delta^2_{i-1}} \left(x^*_{i-1} + \hat{x}_{i-1} \right) $$ 
$$ \Omega_i = \Omega_0 \left( 1- a_i^2 \right) + a_i^2 \Omega_{i-1} \left(1 - \frac{\Omega_{i-1}}{\Omega_{i-1} + \delta^2_{i-1}} \right) $$
$$ a_i = e^{-\left(t_i-t_{i-1}\right)/\tau} $$

To find the optimal parameters we maximize the likelihood with respect to $\sigma_C$ and $\tau$ and calculate $b$ as the mean magnitude of the light curve divided by $\tau$.

\section{The library}
\label{sec:library}

\subsection{Installation}

The library is coded in python and can be downloaded  from the Github repository \url{https://github.com/isadoranun/FATS}. It is also possible to obtain the library by downloading the Python package from \url{http://pypi.python.org/pypi/FATS} or by directly installing it from the terminal as follows:

\begin{verbatim}
 pip install FATS
\end{verbatim}

New features may be added by issuing pull requests via the Github version control system.

\subsection{Input}

The library receives as input the time series data and returns as output an array with the calculated features. Depending on the available input the user can calculate different features. For example, if the user has only the vectors magnitude and time, just the features that need this data will be able to be computed. This will be deeper explained in section \ref{sec:choosing_features}.
In order to calculate all the possible features the following vectors (also termed as raw data) are needed per light curve:
\begin{itemize}
\item magnitude
\item time
\item error
\item magnitude2
\item time2
\item error2
\end{itemize}

where \textit{2} refers to a different observation band.
It is worth pointing out that the \textit{magnitude} vector is the only input strictly required by the library given that it is necessary for the calculation of all the features. The remaining vectors are optional since they are needed just by some features. In other words, if the user does not have this additional data or he is analyzing time series other than light curves, it is still possible to calculate some of the features. More details are presented in  section \ref{sec:choosing_features}.

When observed in different bands, light curves of a same object are not always monitored for the same time length and at the same precise times. For some features, it is important to align the light curves and to only consider the simultaneous measurements from both bands. The \textit{aligned} vectors refer to the arrays obtained by synchronizing the raw data.

Thus, the actual input needed by the library is an array containing the following vectors and in the following order:
\begin{itemize}
\item magnitude
\item time
\item error
\item magnitude2 
\item aligned magnitude
\item aligned magnitude2
\item aligned time
\item aligned error
\item aligned error2
\end{itemize}

To exemplify how the input should be, we provide a basic toolbox for importing and preprocessing the data from the MACHO survey. The function \verb|ReadLC_MACHO| receives a MACHO id of a light curve as an input, processes the file and returns the following output:
\begin{itemize}
\item \textbf{mag}: magnitude measurement 
\item \textbf{time}: time of measurement 
\item \textbf{error}: associated observational error 
\end{itemize}

Besides \verb|ReadLC_MACHO|, the toolbox also provides the following utility functions:
\begin{itemize}
\item  \verb|Preprocess_LC|: points within 5 standard deviations from the mean are considered as noise and thus are eliminated from the light curve. 
\item \verb|Align_LC| : it synchronizes the light curves in the two different bands and  returns \verb|aligned_mag|, \verb|aligned_mag2|, \verb|aligned_time|, \verb|aligned_error| and \verb|aligned_error2|.
\end{itemize}

Note: \verb|mag|, \verb|time|, and \verb|error| must have the same length, as well as \verb|aligned_mag|, \verb|aligned_mag2|, \verb|aligned_time|, \verb|aligned_error|, and \verb|aligned_error2|.\\

The following code is an example of how to use the reading MACHO toolbox:

\begin{python}
#Open the light curve for two different bands
lc_B = FATS.ReadLC_MACHO('lc_58.6272.729.B.mjd')   
lc_R = FATS.ReadLC_MACHO('lc_58.6272.729.R.mjd')

#Import the data
[mag, time, error] = lc_B.ReadLC()
[mag2, time2, error2] = lc_R.ReadLC()

#Preprocess the data
preprocessed_data = FATS.Preprocess_LC(mag, time, error)
[mag, time, error] = preprocessed_data.Preprocess()

preprocessed_data = FATS.Preprocess_LC(mag2, time2, error2)
[mag2, time2, error2] = preprocessed_data.Preprocess()

#Synchronize the data
if len(mag) != len(mag2):
    [aligned_mag, aligned_mag2, aligned_time, aligned_error, aligned_error2] = FATS.Align_LC(time, time2, mag, mag2, error, error2)
    
lc = np.array([mag, time, error, mag2, aligned_mag, aligned_mag2, aligned_time, aligned_error, aligned_error2])
\end{python}


\subsection{Structure}

The library structure is divided into two main parts. Part one: \textbf{Feature.py}, is a wrapper class that allows the user to select the features to be calculated based on the available time series vectors or to specify a list of features. Part two: \textbf{FeatureFunciontLib.py}, contains the actual code for calculating the features. Each feature has its own class with at least two functions:

\textbf{init}: receives the necessary parameters (when needed) for the feature calculation. It also defines the required vectors for the computation (e.g. magnitude and time). 

\textbf{fit}: returns the calculated feature. The output can only take one value; features like the autocorrelation function must consequently be summarized in one single scalar.

The following code is an example of a class in \textbf{FeatureFunciontLib.py} that calculates the slope of a linear fit to the light curve:\\

\begin{python}
class LinearTrend(Base):

    def __init__(self):
     
    	self.Data=['magnitude','time']

    def fit(self,data):
        
       	magnitude = data[0]
        
        time = data[1]
        
        regression_slope = stats.linregress(
        time, magnitude)[0]

        return regression_slope
\end{python}

If the user wants to contribute with features to the library, they must be added to \textbf{FeatureFunctionLib.py} following the explained format. There is no need to modify \textbf{Feature.py}. The user should also add a validity test to the unit test file (see explanation in section \ref{sec:unit_test} ).

\subsection{Choosing the features to calculate}
\label{sec:choosing_features}

The library  allows the user to either choose the specific features of interest to be calculated, or to calculate them all simultaneously. Nevertheless, as already mentioned, the features are divided depending on the input data needed for their computation (magnitude, time, error, second data, etc.). If unspecified, this will be used as an automatic selection parameter. For example, if the user wants to calculate all the available features but only has the vectors \textit{magnitude} and \textit{time}, only the features that need \textit{magnitude} and/or \textit{time} as an input will be computed.\\

\textbf{Note:} some features depend on other features and consequently must be computed together. For instance, \textit{Period\_fit} returns the false alarm probability of the estimated period. It is necessary then to calculate also the period \textit{PeriodLS}.\\

The list of all the possible features with their corresponding input data, additional parameters and literature source is presented in Table \ref{table:features}.

\begin{table*}[h!]
\begin{center}
\resizebox{\textwidth}{!}{%
 \begin{tabular}{||c | c |c| c| c| c||} 
 \hline
Feature & Depends on & Input data (besides magnitude) & Parameters & Default & Ref \\ [0.5ex] 
 \hline\hline
 Amplitude &  &  &  &  & \citet{Richards2011} \\  
 \hline
 Anderson\-Darling test  &  & &  &  & \citet{Kim2008} \\
 \hline
 Autocor\_length &  &  & Number of lags & 100 & \citet{Kim2011} \\
 \hline
 Beyond1Std &  & error  &  &  & \citet{Richards2011} \\
  \hline
 CAR\_mean  & CAR\_sigma  & time, error  &  &  & \citet{Pichara2012} \\
  \hline
CAR\_sigma  &  & time, error  & &  &  \citet{Pichara2012} \\
  \hline
CAR\_tau  & CAR\_sigma & time, error   &  &  & \citet{Pichara2012} \\
  \hline
Color &  & mag2 & &  &   \citet{Kim2011} \\
  \hline
Con  &  &  &  Consecutive stars & 3 & \citet{Kim2011} \\
  \hline
Eta\_color &  & aligned\_mag, aligned\_mag2, aligned\_time & &    & \citet{Kim2014} \\
  \hline
Eta\_e  &  & time  & &  &   \citet{Kim2014} \\
  \hline
FluxPercentileRatioMid20    &  & &  &  & \citet{Richards2011} \\
  \hline
FluxPercentileRatioMid35    &  & &  &  & \citet{Richards2011} \\
  \hline
FluxPercentileRatioMid50    &  & &  &  & \citet{Richards2011} \\
  \hline
FluxPercentileRatioMid65    &  & &  &  & \citet{Richards2011} \\
  \hline
FluxPercentileRatioMid20    &  & &  &  & \citet{Richards2011} \\
  \hline
FluxPercentileRatioMid80    &  & &  &  & \citet{Richards2011} \\
  \hline
Freq1\_harmonics\_amplitude\_0  &    & time &  &  & \citet{Richards2011} \\
 \hline
 Freq\{i\}\_harmonics\_amplitude\_\{j\}	  &    & time &  &  & \citet{Richards2011} \\
 \hline
Freq\{i\}\_harmonics\_rel\_phase\_\{j\}	  &    & time &  &  & \citet{Richards2011} \\
 \hline
 LinearTrend  &    & time &  &  & \citet{Richards2011} \\
 \hline
 MaxSlope  &    & time &  &  & \citet{Richards2011} \\
 \hline
 Mean  &    & &  &  & \citet{Kim2014} \\
 \hline
 Meanvariance  &    & &  &  & \citet{Kim2011} \\
 \hline
MedianAbsDev  &    & &  &  & \citet{Richards2011} \\
 \hline
 MedianBRP  &    & &  &  & \citet{Richards2011} \\
 \hline
PairSlopeTrend &    & &  &  & \citet{Richards2011} \\
 \hline
PercentAmplitude  &    & &  &  & \citet{Richards2011} \\
 \hline
PercentDifferenceFluxPercentile	  &    & &  &  & \citet{Richards2011} \\
 \hline
PeriodLS  &    &time &Oversampling factor  & 6 & \citet{Kim2011} \\
 \hline
Period\_fit  &    &time &  &  & \citet{Kim2011} \\
 \hline
Psi\_CS  &    &time &  &  & \citet{Kim2014} \\
 \hline
Psi\_eta  &    &time &  &  & \citet{Kim2014} \\
 \hline
 Q31  &    & &  &  & \citet{Kim2014} \\
 \hline
Q31\_color  &    &aligned\_mag, aligned\_mag2 &  &  & \citet{Kim2014} \\
 \hline
 Rcs  &    & &  &  & \citet{Kim2011} \\
 \hline
Skew  &    & &  &  & \citet{Richards2011} \\
 \hline
SlottedA\_length &    & time&Slot size T (days)  & 4  & \citet{protopapas2015} \\
 \hline
 SmallKurtosis  &    & &  &  & \citet{Richards2011} \\
 \hline
Std  &    & &  &  & \citet{Richards2011} \\
 \hline
StetsonJ &    &aligned\_mag, aligned\_mag2, aligned\_error, aligned\_error2 &  &  & \citet{Richards2011} \\
 \hline
StetsonK  &    & error&  &  & \citet{Richards2011} \\
 \hline
StetsonK\_AC  &    & &  &  & \citet{Kim2011} \\
 \hline
StetsonL  &    & aligned\_mag, aligned\_mag2, aligned\_error, aligned\_error2&  &  & \citet{Kim2011} \\
 \hline
VariabilityIndex  &    & &  &  & \citet{Kim2011} \\ [1ex] 
 \hline
\end{tabular}}
\caption{List of features}
\label{table:features}
\end{center}
\end{table*}

The possible ways of how an user can choose the features from the library to be calculated are presented next.

\subsubsection{List of features as input}

The user can also specify a list of features as input by specifying the features as a list for the parameter \verb|featureList|. In the following example, we aim to calculate the standard deviation and Stetson L of the data:

\begin{python}
a = FATS.FeatureSpace(featureList=['Std', 'StetsonL'])
a = a.calculateFeature(lc)
\end{python}

\subsubsection{Available data as input}

In case the user does not have all the input vectors mentioned above, it is necessary to specify the available  data by specifying the list of vectors using the parameter \verb|Data|. In the example below, we calculate all the features that can be computed with the \textit{magnitude} and  \textit{time} as an input.

\begin{python}
a = FATS.FeatureSpace(Data=['magnitude', 'time'])
a = a.calculateFeature(lc)
\end{python}

\subsubsection{List of features and available data as input}

It is also possible to provide the available time series input vectors and calculate all possible features from a feature list using this data:\\

\begin{python}
a = FATS.FeatureSpace(featureList=['Mean', 'Beyond1Std', 'CAR_sigma','Color', 'SlottedA_length'], Data = ['magnitude', 'error'])
a = a.calculateFeature(lc)}
\end{python}

In this case only the features \textit{Mean} and \textit{Beyond1Std} could be computed given the data, and warnings would be thrown for the other features:\\

\textit{Warning: the feature CAR\_sigma could not be calculated because [`magnitude', `time', `error'] are needed.}\\

\textit{Warning: the feature Color could not be calculated because [`magnitude', `time', `magnitude2'] are needed.}\\

\textit{Warning: the feature SlottedA\_length could not be calculated because [`magnitude', `time'] are needed.}

\subsubsection{Excluding list as input}

The user can also create a list of features to be excluded from the calculation. To do so, the list of features to be excluded can be passed as a list via the parameter \verb|excludeList|. For example:\\

\begin{python}
a = FATS.FeatureSpace(Data=['magnitude', 'time', 'error'], excludeList=['SlottedA_length', 'StetsonK_AC', 'PeriodLS'])
a = a.calculateFeature(lc)
\end{python}

The following list of features would be calculated: 
\textit{Amplitude,
 AndersonDarling,
 Autocor\_length,
 Beyond1Std,
 CAR\_mean,
 CAR\_sigma,
 CAR\_tau,
 Con,
 Eta\_e,
 FluxPercentileRatioMid features,
 Harmonics features,
 LinearTrend,
 MaxSlope,
 Mean,
 Meanvariance,
 MedianAbsDev,
 MedianBRP,
 PairSlopeTrend,
 PercentAmplitude,
 PercentDifferenceFluxPercentile,
 Period\_fit,
 Psi\_CS,
 Psi\_eta,
 Q31,
 Rcs,
 Skew,
 SmallKurtosis,
 Std,
 and StetsonK}

  and the following warnings would be displayed:

\textit{Warning: the feature Color could not be calculated because [`magnitude', `time', `magnitude2'] are needed.}\\

\textit{Warning: the feature Eta\_color could not be calculated because [`magnitude', `time', `magnitude2'] are needed.}\\

\textit{Warning: the feature Q31\_color could not be calculated because [`magnitude', `time', `magnitude2'] are needed.}\\

\textit{Warning: the feature StetsonJ could not be calculated because [`magnitude', `time', `error', `magnitude2', `error2'] are needed.}\\

\textit{Warning: the feature StetsonL could not be calculated because [`magnitude', `time', `error', `magnitude2', `error2'] are needed.}\\

\textit{error: please run PeriodLS first to generate values for Period\_fit}\\

\textit{error: please run PeriodLS first to generate values for Psi\_CS}\\

\textit{error: please run PeriodLS first to generate values for Psi\_eta}\\

\subsubsection{All the features}

To calculate all the existing features in the library, the user can set the \verb|Data| parameter to the string \verb|'all'| and set the \verb|featureList| parameter to be \verb|None|. Obviously, in order to do this, the user must have also provided all the necessary time series input vectors.

\subsection{Library output}

When calculating the features of a light curve, the output can be returned in three different formats:
\begin{itemize}
\item \textit{dict}: returns a vector with the name of each feature followed by its value.
\item \textit{array}: returns a vector with only the features values.
\item \textit{features} : returns a vector with the list of the calculated features.
\end{itemize}

The output format can be specified via a string for the parameter \verb|method|, as shown in the example below:

\begin{python}
a = FATS.FeatureSpace(Data='all', featureList=None) 
a = a.calculateFeature(lc)
print a.result(method='dict')
\end{python}

\section{Tests}

\subsection{Robustness tests}
\label{sec:invariance}

The following section presents the tests performed to the features in order to check their invariance to unequal sampling \footnote{The code is also available in the github repository}. To do so, we took random observations of a light curve and compared the resulting features with the ones obtained from the original data. The steps we followed are described next:
\begin{itemize}
\item we calculate the features for fifty random samples of one original light curve.
\item we obtain the mean and standard deviation of each calculated feature.
\item we calculate the percentage difference between the original features' values and the ones obtained from the random sampling.
\end{itemize}

\subsection{Unit tests}
\label{sec:unit_test}

In order to systematically check the correctness of the feature generation implementation in our library, a unit test is created for each feature in a unit test file named \verb|test_library.py|. This script should be always run before using the library by executing \verb|$ py.test| at the command line. In addition, if a user contributes a new feature for the library, a pertinent test should be added to the unit test file. The idea is to guarantee, as far as possible, that every feature calculated is correct. In most cases, this can be reached by calculating the expected feature value for a known distribution (normal, uniform, or otherwise), and then checking it with the value obtained from the library.

The following image shows how \verb|py.test| output should look if all the tests are passed:

\begin{figure}[H]
\centering
\includegraphics[width=8.5cm]{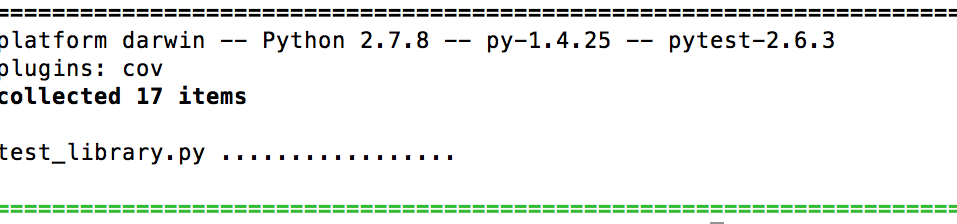}
\caption{Unit test output.}
\label{fig:curvas_ejemplos}
\end{figure}

\subsection{Classification test}

As a way of checking the usefulness of our features we trained a classifier and obtained the classification accuracy for labeled objects that were not used in the training set. In particular we used a Random Forest classifier and calculated the F-score of the out-of-bag (oob) samples as done in \citet{Nun2014a}. Since trees are constructed from different bootstrapped samples of the original data, about one-third of the cases are left out of the \textquotedblleft bag\textquotedblright  and not used in the  construction of each tree.  By putting  these oob observations down the trees that were not trained with oob data,  we end up with unbiased predicted labels. The training set we used for the test is composed of a subset of 6063 labeled observations and 8 different classes from the MACHO catalog \citep{Kim2011}\footnote{We collected these variables from the MACHO variable catalog found at: http://vizier.u-strasbg.fr/viz-bin/VizieR?-source=II/247}. The constitution of the training set is presented in Table \ref{Table:Training}:

\begin{table}[H]
\caption{\label{Table:Training}Training Set Composition}
\centering
\begin{tabular}{ c | c | c |}  
\cline{2-3}
{} & Class & Number of objects \\
\hline
\multicolumn{1}{ |c| }{1} & Be Stars  & $127$ \\
\multicolumn{1}{ |c| }{2} & Cepheid  & $101$ \\
\multicolumn{1}{ |c| }{3} & Eclipsing Binaries & $255$ \\
\multicolumn{1}{ |c| }{4} & Long Period Variable & $365$ \\
\multicolumn{1}{ |c| }{5} & MicroLensing  & $380$ \\
\multicolumn{1}{ |c| }{6} & Non variables  & $3966$ \\ 
\multicolumn{1}{ |c| }{7} & Quasars  & $59$ \\
\multicolumn{1}{ |c| }{8} & RR Lyrae & $610$ \\
\hline
\end{tabular}
\end{table}

By using 500 trees we obtain a F-score of 0.97 and the following confusing matrix:

\begin{figure}[H]
\centering
\includegraphics[width=8.5cm]{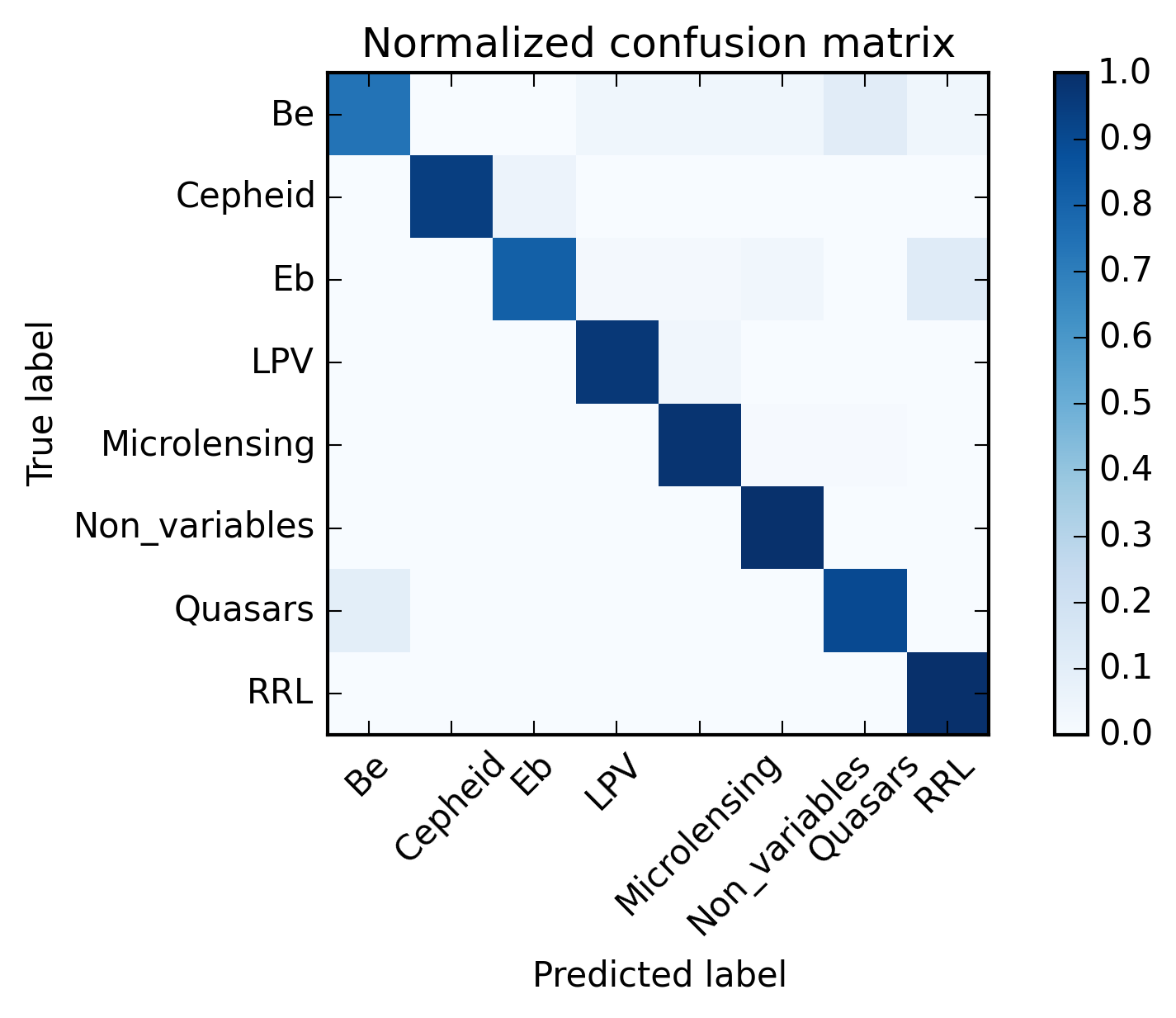}
\caption{Confusion matrix for the MACHO training set}
\label{fig:confusion_matrix}
\end{figure}

\begin{figure*}[h!]
\centering
\includegraphics[width=14cm]{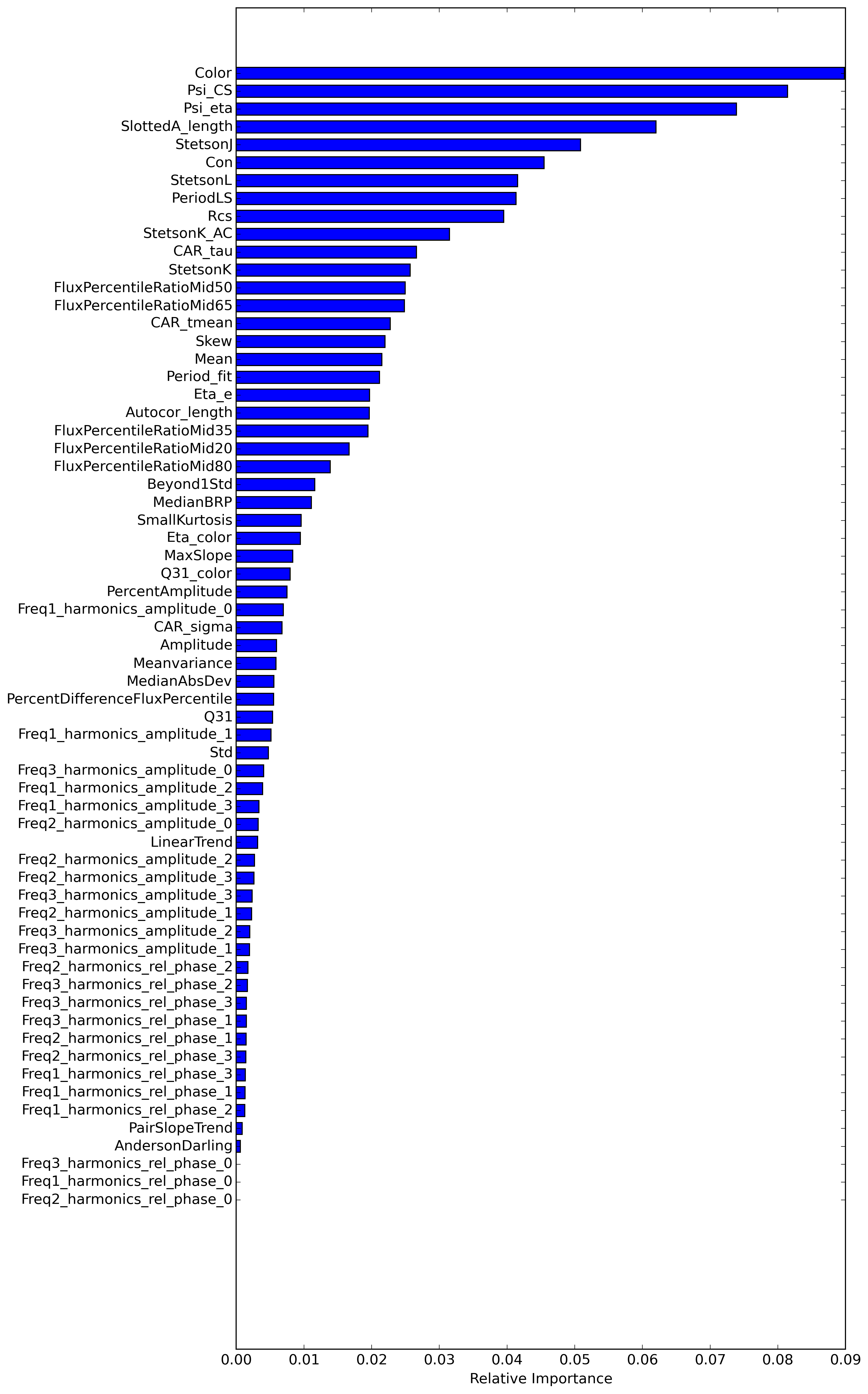}
\caption{Feature importance}
\label{fig:importance}
\end{figure*}

We also obtained the importance ranking of the features in the classification task as shown in  figure \ref{fig:importance}. To understand how the importance is calculated the reader can find information in \citet{Breiman2001}.


\section{How to contribute}

A user can contribute to this work in two different ways:
\begin{itemize}
\item by adding a new feature to the library with its pertinent description, validity test and unit test.
\item by further developing or correcting an already existing feature.
\end{itemize}

In the first case, the user can directly download the github package of the library and push a new version with the added feature. After being checked, it will be officially uploaded and the user will be added as a co-author to this arXiv article.
In the second case, the user can contact us by email with any suggestions.

\section{Upcoming features}	

The latest release of the package is FATS 1.3.4, uploaded on April 8, 2015. The next features to be added are:
\begin{itemize}
\item structure function descriptor
\item slepian wavelet variance \citep{Graham2015}
\end{itemize}

\section{Summary}
\label{sec:conclusions}
In this article we presented FATS, a python library for times series feature extraction. In particular we focused on astronomical times series (light-curves), nevertheless the tool is not restricted to this type of data. We included the main descriptors for times series in a simple collaborative package. Several tests were performed to check their robustness and invariance to nonuniform sampling. We also proved the suitability of the features for classification by obtaining a high F-score of 0.97 with a Random-forest classifier and data from the MACHO catalog.
FATS is under continuous development and features will be constantly added. The author list will be as well updated insofar as users send us their contribution.

\section{Acknowledgments}
We would like to thank Adolfo Hurel, And\'es Riveros and Javier Machin for their useful discussions and comments.

\clearpage
\bibliographystyle{apj}
\bibliography{FATS}

\begin{thebibliography}{}
\expandafter\ifx\csname natexlab\endcsname\relax\def\natexlab#1{#1}\fi

\bibitem[{Breiman(2001)}]{Breiman2001}
Breiman, L. 2001, Machine learning, 45, 5

\bibitem[{Brockwell(2002)}]{brockwell2002}
Brockwell, P.~J. 2002, Introduction to time series and forecasting, Vol.~1
  (Taylor \& Francis)

\bibitem[{Debosscher {et~al.}(2007)Debosscher, Sarro, Aerts, Cuypers,
  Vandenbussche, Garrido, \& Solano}]{debosscher2007}
Debosscher, J., Sarro, L., Aerts, C., {et~al.} 2007, Astronomy \& Astrophysics,
  475, 1159

\bibitem[{Ellaway(1978)}]{Ellaway1978}
Ellaway, P. 1978, Electroencephalography and Clinical Neurophysiology, 45, 302

\bibitem[{Falk(2012)}]{Falk2012}
Falk, M. 2012, {A First Course on Time Series Analysis}

\bibitem[{Graham(2015)}]{Graham2015}
Graham, M.~J. 2015, in Tools for Astronomical Big Data Tucson, Arizona, March
  9-11, 2015 Talk Abstracts

\bibitem[{Huijse {et~al.}(2012)Huijse, Estevez, Protopapas, Zegers, \&
  Principe}]{huijse2012}
Huijse, P., Estevez, P.~A., Protopapas, P., Zegers, P., \& Principe, J.~C.
  2012, Signal Processing, IEEE Transactions on, 60, 5135

\bibitem[{Kim {et~al.}(2008)Kim, Protopapas, Alcock, Byun, \& Bianco}]{Kim2008}
Kim, D.-W., Protopapas, P., Alcock, C., Byun, Y.-I., \& Bianco, F. 2008, 13, 1

\bibitem[{Kim {et~al.}(2014)Kim, Protopapas, Bailer-Jones, Byun, Chang,
  Marquette, \& Shin}]{Kim2014}
Kim, D.-W., Protopapas, P., Bailer-Jones, C. a.~L., {et~al.} 2014, 18

\bibitem[{Kim {et~al.}(2011)Kim, Protopapas, Byun, Alcock, Khardon, \&
  Trichas}]{Kim2011}
Kim, D.-W., Protopapas, P., Byun, Y.-I., {et~al.} 2011, The Astrophysical
  Journal, 735, 68

\bibitem[{Nun {et~al.}(2014)Nun, Pichara, Protopapas, \& Kim}]{Nun2014a}
Nun, I., Pichara, K., Protopapas, P., \& Kim, D.-W. 2014, 17

\bibitem[{Pichara {et~al.}(2012)Pichara, Protopapas, Kim, Marquette, \&
  Tisserand}]{Pichara2012}
Pichara, K., Protopapas, P., Kim, D., Marquette, J., \& Tisserand, P. 2012, 12,
  1

\bibitem[{{Protopapas} {et~al.}(2015){Protopapas}, {Huijse}, {Est{\'e}vez},
  {Zegers}, {Pr{\'{\i}}ncipe}, \& {Marquette}}]{protopapas2015}
{Protopapas}, P., {Huijse}, P., {Est{\'e}vez}, P.~A., {et~al.} 2015, \apjs,
  216, 25

\bibitem[{Richards {et~al.}(2011)Richards, Starr, Butler, Bloom, Brewer,
  Crellin-Quick, Higgins, Kennedy, \& Rischard}]{Richards2011}
Richards, J.~W., Starr, D.~L., Butler, N.~R., {et~al.} 2011, The Astrophysical
  Journal, 733, 10

\bibitem[{Scargle(1982)}]{scargle1982}
Scargle, J.~D. 1982, The Astrophysical Journal, 263, 835

\bibitem[{Stetson(1996)}]{Stetson1996}
Stetson, P.~B. 1996, Publications of the Astronomical Society of the Pacific,
  108, pp. 851

\bibitem[{Von~Neumann {et~al.}(1941)Von~Neumann, Kent, Bellinson, \&
  Hart}]{von1941}
Von~Neumann, J., Kent, R., Bellinson, H., \& Hart, B.~t. 1941, The Annals of
  Mathematical Statistics, 12, 153

\bibitem[{Wozniak(2000)}]{wozniak2000}
Wozniak, P. 2000, arXiv preprint astro-ph/0012143

\end{thebibliography}

\end{document}